\documentclass[12pt]{iopart}
\usepackage{bm}
\usepackage{color}
\usepackage{amssymb}
\usepackage{subfigure}
\newcommand{\be}{\begin{equation}}
\newcommand{\ee}{\end{equation}}\newcommand{\bea}{\begin{eqnarray}}
\newcommand{\eea}{\end{eqnarray}}

\usepackage{graphicx}
\usepackage{color}
\usepackage{dcolumn}
\usepackage{bm}

\begin{document}

\title[Nonadiabatic generation of spin currents in a quantum ring]  
{Nonadiabatic generation of spin currents in a quantum ring with    
Rashba and Dresselhaus spin-orbit interactions}

\author{Marian Ni\c t\u a$^{1}$, D.~C.~ Marinescu$^{2}$, Bogdan Ostahie$^{1}$,
Andrei Manolescu$^{3,*}$, Vidar Gudmundsson$^{4}$}
\address{$^{1}$National Institute of Materials Physics, P.O. Box MG-7,
Bucharest-Magurele, Romania}
\address{$^{2}$Department of Physics and Astronomy, Clemson University, Clemson,
South Carolina 29634, USA}
\address{$^{3}$School of Science and Engineering, Reykjavik University, Menntavegur 1,
IS-101 Reykjavik, Iceland}
\address{$^{4}$Science Institute, University of Iceland, Dunhaga 3, IS-107 Reykjavik,
Iceland}
\address{$^{*}$ E-mail: manoles@ru.is}



\begin{abstract}
When subjected to a linearly polarized terahertz pulse, a mesoscopic ring
endowed with spin-orbit interaction (SOI) of the Rashba-Dresselhaus type
exhibits non-uniform azimuthal charge and spin distributions. Both types of 
SOI couplings are considered linear in the electron momentum.  Our results are
obtained within a formalism based on the equation of motion satisfied by
the density operator which is solved numerically for different values of
the angle $\phi$, the angle determining the polarization direction
of the laser pulse.  Solutions thus obtained are later employed in
determining the time-dependent charge and spin currents, whose values
are calculated in the stationary limit.  Both these currents exhibit
an oscillatory behavior complicated in the case of the spin current
by a beating pattern.  We explain this occurrence on account of the two
spin-orbit interactions which force the electron spin to oscillate between
the two spin quantization axes corresponding to Rashba and Dresselhaus
interactions.  The oscillation frequencies are explained using the
single particle spectrum.
\end{abstract}


\maketitle

\section{Introduction}

The controlled induction of spin and charge currents in mesoscopic rings
represents a longstanding interest of experimentalists and theorists
alike. By taking advantage of the right-left asymmetry of the electron
states, various methods have been proposed over the years to this end
\cite{moskalets1,moskalets2, gudmundsson}. In particular, by applying
an ultrashort, terahertz frequency laser pulse with spatial asymmetry,
the charge and spin excitations are realized in a non-adiabatic regime,
since their excitation occurs on a time scale that is much shorter
than the electron relaxation lifetime \cite{gudmundsson,siga,nita}. More
recently, these ideas have been reconsidered in the case of rings made out
of zinc-blend semiconductor structures that exhibit spin-orbit interaction
(SOI). The linear coupling between the electron momentum and spin
appears as a result of broken inversion symmetry that results from spatial
confinement (Rashba - R) \cite{rashba} or from bulk properties (Dresselhaus
- D) \cite{dresselhaus}.  Under these general circumstances, it was shown
that the spin and charge flow simultaneously \cite{splettstoesser, souma,
sheng} and special configurations, such as hybrid structures \cite{sun}
or a number constraint (odd number of particles) \cite{huang} are needed
to isolate the spin contribution. Recently, we demonstrated that by applying
a two-component terahertz frequency pulse, for certain values of
the phase difference angle $\phi$ between the two components, a pure spin current
is non-adiabatically induced in the ring. Physically, this situation is
a consequence of the interplay between SOI which rotates the electron
spin around the ring and the asymmetry of the external perturbation
embodied by $\phi$.

The present paper discusses a follow-up to this analysis by studying
the effect of a terahertz laser pulse on a ring endowed with both
Rashba and Dresselhaus spin-orbit couplings. The coexistence of these
couplings introduces precession effects of the electron spin along
different directions, a phenomenon bound to affect the spin and charge
distributions around the ring. Our formalism is based on obtaining
the stationary solutions of the equation of motion satisfied by the
particle-density operator which are later employed in estimating the spin
and charge currents as as statistical averages. The three independent
parameters of the problem are the phase of the laser excitation, $\phi$,
and the coupling strengths of the Rashba and Dresselhaus interactions.
Our main result is that, for comparable R and D intensities, the spin
current, calculated along the $\hat{z}$ direction,
showcases an oscillatory pattern that results from the nutation of the
electron spin between the two natural directions associated with the
two spin-orbit interactions.

\section{The Equilibrium Hamiltonian}

We consider a one-dimensional (1D) quantum ring of radius $r_0$ in the
presence of both Rashba and Dresselhaus spin-orbit interactions.  The ring
is discretized in $N$ sites with a lattice constant $a=2\pi r_0/N$.
The polar coordinates are chosen such that the site with index $n$ has an
azimuthal angle $\theta_n=2n\pi/N$.  We denote by $c^{+}_{n,\sigma}$ and
$c_{n,\sigma}$ the creation and annihilation operators for an electron
with spin $\sigma$ on the site $n=1,...,N$.  

In the following considerations we adopt the discrete representation of the Hamiltonian $
H_{\rm ring}=H_0+V_{R}+V_{D} \,,
$ that describes the non-interacting electron dynamics around the ring in the presence of the Rashba (R) and Dresselhaus (D) SOI couplings. Its components are:
\begin{equation}\label{h0}
H_0 = 2V\sum_{n,\sigma}c^{+}_{n,\sigma}c^{}_{n,\sigma}
-V\sum_{n,\sigma} c^{+}_{n,\sigma}c^{}_{n+1,\sigma}
-V\sum_{n,\sigma} c^{+}_{n,\sigma}c^{}_{n-1,\sigma},
\end{equation}
\begin{eqnarray}\label{vr}
      V_{R}=&&
      -iV_{\alpha}\sum_{n,\sigma,\sigma'}
      {\bf \sigma}_r(\theta_{n,n+1})^{\sigma, \sigma'}
      c^{+}_{n \sigma} c^{}_{n+1 \sigma'}+{\rm h.c.}\, ,
\end{eqnarray}
\begin{eqnarray}\label{vd}
      V_{D}=&&
      -iV_{\beta}\sum_{n,\sigma,\sigma'}
      {\bf \sigma}_\theta^*(\theta_{n,n+1})^{\sigma, \sigma'}
      c^{+}_{n \sigma} c^{}_{n+1 \sigma'}+{\rm h.c.} \, .
\end{eqnarray}
Three different energy scales are introduced: the hopping energy $V=\hbar^2/2m^*a^2$ , Rashba energy $V_{\alpha}=\alpha/2a$ and Dresselhaus energy $V_{\beta}=\beta/2a$,
with $\alpha$ and $\beta$ being SOI parameters and $m^*$  the effective mass of the electron in the
host semiconductor material.
 It is customary to introduce the radial and azimuthal spin matrices
$\sigma_r$ and $\sigma_\theta$ 
\begin{figure}
\begin{center}
\includegraphics[width=6cm]{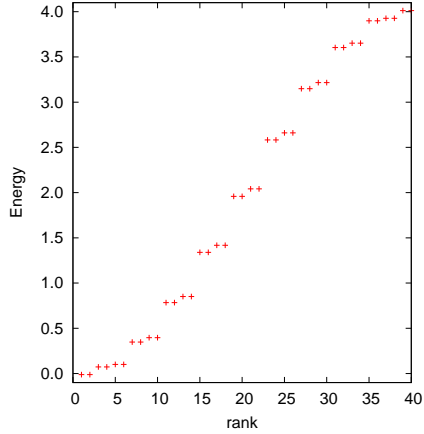}
\end{center}
\caption{
The energy spectrum of a ring with SOI. The Rashba energy $V_\alpha=0.1$ and 
the Dresselhaus energy $V_\beta=0.05$.  On the horizontal axis is the rank of the
energy value within the spectrum of the Hamiltonian. All states are spin degenerate.
The energy unit is $V=\hbar^2/2m^*a^2$.  At the initial time only the first 6 energy
levels are occupied.
}
\label{spectru}
\end{figure}
\begin{equation}
{\bf \sigma}_r(\theta)={\bf \sigma}_x \cos\theta+{\bf \sigma}_y \sin\theta \, , \,\,\,
{\bf \sigma}_\theta(\theta)=-{\bf\sigma}_x \sin\theta+ {\bf \sigma}_y \cos\theta\, .
\end{equation}
In the presence of only one type of SOI, either Rashba or Dresselhaus, the eigenstates
of the Hamiltonian can be obtained analytically \cite{souma,sheng,splettstoesser}.
In both cases, the spin quantization axis is tilted relatively to 
the $z$ axis by angles $\Omega_{\alpha}$ and $\Omega_{\beta}$
respectively, given by \begin{equation}
\tan \Omega_{\alpha,\beta}=\frac{V_{\alpha,\beta}}{V\sin(\Delta\theta/2)} \,\, ,
\end{equation}
where $\Delta\theta=2\pi/N$.  Consequently, if either $V_{\beta}=0$
or $V_{\alpha}=0$, there exist a corresponding natural direction of the
spin, given by 
${\bf e}_{\alpha}=\cos \Omega_\alpha {\bf e}_z -\sin \Omega_\alpha {\bf e}_r$ 
or 
${\bf e}_{\beta }=\cos \Omega_\beta {\bf e}_z +\sin \Omega_\beta {\bf e}_{-\theta}$, 
respectively, along which the
projection of the electron spin is conserved and equal to $\pm \hbar/2$.
The directions ${\bf e}_{\alpha }$ and ${\bf e}_{\beta }$ are called the
principal spin axes, or Rashba and Dresselhaus spin quantization axes,
respectively.  The difference between the Rashba and the Dresselhaus
SOIs is reflected by the orientation of the spin precession relatively to
the orbit:  in the Rashba case, the precession and the orbital angular
speed have the same sign, whereas in the Dresselhaus case they are
opposite \cite{Ihn,sheng}.

When both interactions are present the diagonalization of the Hamiltonian
can be performed only by numerical methods.  The energy spectrum for
$N = 20$ sites and with $V_{\alpha}=0.1 V$ and $V_{\beta}=0.05 V$ is
shown in Fig.~\ref{spectru}.  Each state is double (spin) degenerated.
For a ring radius $r_0=50$ nm our SOI strengths become comparable to
those in experimental materials \cite{Ihn}.  One obtains $\alpha\approx
21, \ \beta\approx 10$ meVnm for InAs (i.\ e.\ using $m^*=0.023m_e$),
or $\alpha\approx 34, \ \beta\approx 17$ meVnm for InSb (i.\ e.\ using
$m^*=0.014m_e$).  Our time unit is $t=\hbar/V$, corresponding to $\approx 0.1$ ps for
InAs and $\approx 0.06$ ps for InSb.
%
%
We will consider $n_e = 6$ (noninteracting) electrons in the ring.
The occupied states have energies
$E_{1,2}=-0.01184, E_{3,4}=0.07311, E_{5,6}= 0.1003$.  The next
three energy values are $E_{7,8}= 0.3484, E_{9,10}=0.3967, \ \rm {and}
\ E_{11,12}=0.3967$ (units of $V$).  The even and odd labels correspond
to positive and negative spin chiralities, respectively.

\begin{figure}
\begin{center}
\includegraphics[width=10cm]{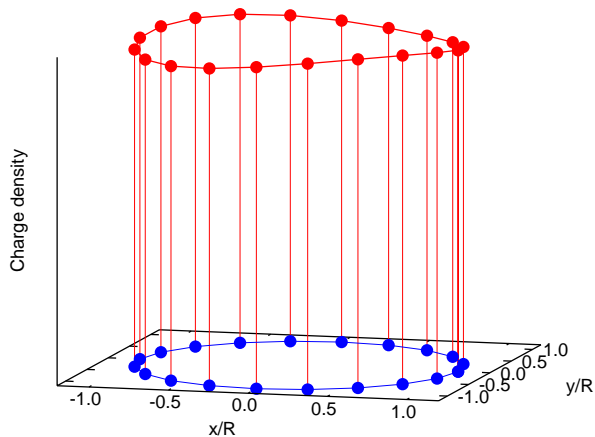}
\end{center}
\vspace{-1.5cm}
\caption{
The electron density around the ring.  Due to the presence of both
SOIs the distribution of electrons is nonuniform.  Minima occur along
the direction $y=x$, or at angles $\theta=\pi/4$ and $\theta=5\pi/4$,
and maxima along $y=-x$ or at angles $\theta=3\pi/4$ and $\theta=7
\pi/4$. The blue dots indicate the $N=20$ sites on the ring.
}
\label{CDW}
\end{figure}

The combination of the two types of SOI breaks the circular symmetry of the
electron distribution and creates a charge deformation as shown in Fig.\ \ref{CDW}.
This effect has been discussed in the recent literature both for one-dimensional
\cite{sheng} and two-dimensional \cite{nowak} quantum rings. In the
one-dimensional ring model the charge density has two minima along the direction
$y=x$ and two maxima along $y=-x$.
The effect can be explained by an intrinsic periodic potential that
develops around the ring \cite{sheng},
\begin{equation}
U(\theta)=2 V_{\alpha} V_{\beta} \sin (2\theta) \, .
\label{uper}
\end{equation}

\section{Time dependent perturbation}

At time $t=0$ the quantum ring is excited by a short terahertz laser pulse
linearly polarized along a direction making the angle $\phi$ with the
$y$ axis.  The pulse has a lifetime of $1/\Gamma$ and an angular frequency
$\omega$ and can be represented by the time dependent Hamiltonian
\begin{eqnarray}\label{ht1}
H_{\rm pulse}(t)&=&A e^{-\Gamma t} \sin (\omega t)  \cos (\theta + \phi) \ .
\end{eqnarray}
The time evolution of the system is given by the Liouville equation,
\begin{equation}
i\hbar \dot \rho(t) = \left[ H + H_{pulse}(t), \ \rho(t) \right ] \, ,
\label{eq:L}
\end{equation}
where $\rho(t)$ is the time dependent statistical operator. The time evolution of the electron states, assumed to be initially in the minimum energy configuration, is described by the numerical solutions of the Liouville equation obtained
with the Crank-Nicolson finite difference method 
\cite{gudmundsson} with small time steps $\delta t\ll \Gamma^{-1}$.
The expectation value of any observable $O$ is then calculated as $\langle
O\rangle= {\rm Tr} \left[\rho(t)O\right]$.

\section{Numerical results}

\begin{figure}
\centering
\includegraphics[width=14cm]{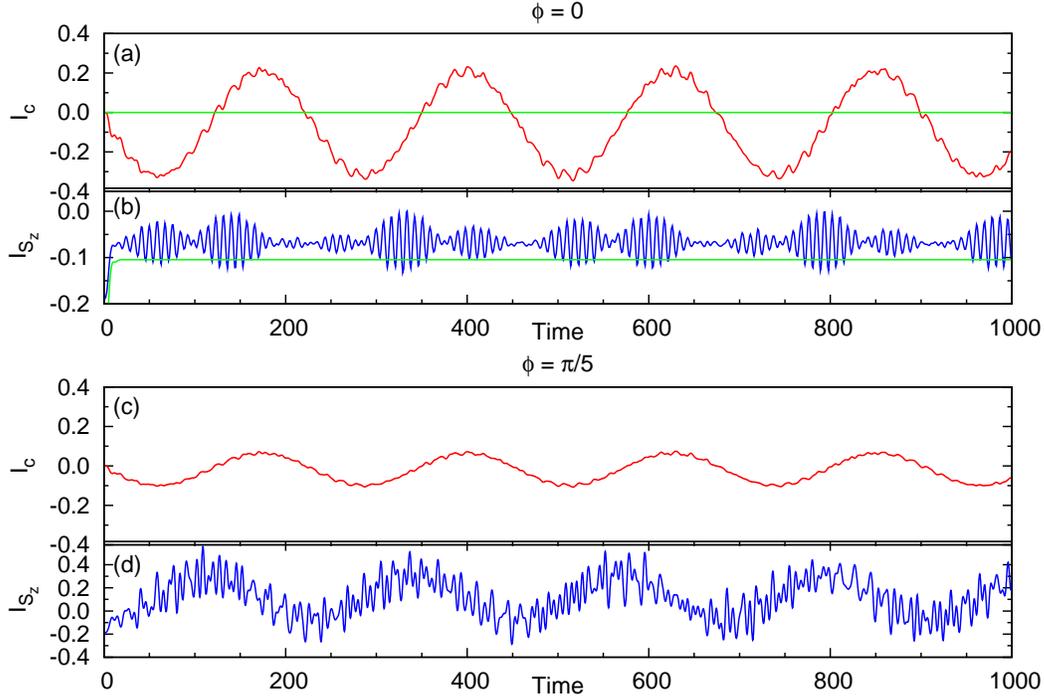}
\caption{
Charge and spin currents in the presence of both Rashba and Dresselhaus SOI.
$V_{\alpha}=0.1V$  and $V_{\beta}=0.05V$. The upper panels show the results for
the relative angle $\phi=0$, and the lower panels for $\phi=\pi/5$. The oscillations
are explained by the Bohr frequencies of the energy spectrum, see text. The green lines
in the upper panels show the results obtained in \cite{nita} with only one type of SOI
interaction.
The time is in units of $\hbar/V$
}
\label{chargeRD}.
\end{figure}

In the following example the pulse parameters are $A=2.3V$, $\omega=0.3
V/\hbar$, and $\Gamma=\omega$, generating a pulse lifetime of
$3.3\hbar/V$.  The time evolution of the charge current $I_c(t)=ev(t)$
for $\phi=0$ is shown in panel (a) of Fig.\ \ref{chargeRD}.  After the
pulse vanishes, the charge current exhibits a large period oscillatory behavior. At high frequencies, superimposed secondary oscillations are noticeable.
These findings can be explained on account of single electron excitations. The large
period is correlated with transitions between the second and the third
states on the energy scale, $\{3,4\}$ and $\{5,6\}$ respectively, that generate
$2\pi/(E_{6}-E_{4})\approx 231\hbar/V$.  Transitions between the states
$\{3,4\}$ and $\{1,2\}$, occur with period $74\hbar/V$, i.\ e.\ one third of the large
period, have less weight and cannot be distinguished from the large oscillations.
Excitations over larger energy intervals, like $2\pi/(E_{8}-E_{6})\approx
23\hbar/V$, or $2\pi/(E_{10}-E_{2})\approx 15\hbar/V$, etc. are
responsible for the finer oscillations, with periods varying from
$50\hbar/V$ down to $15 \hbar/V$ or less.

\begin{figure}
\centering
\includegraphics[width=15cm]{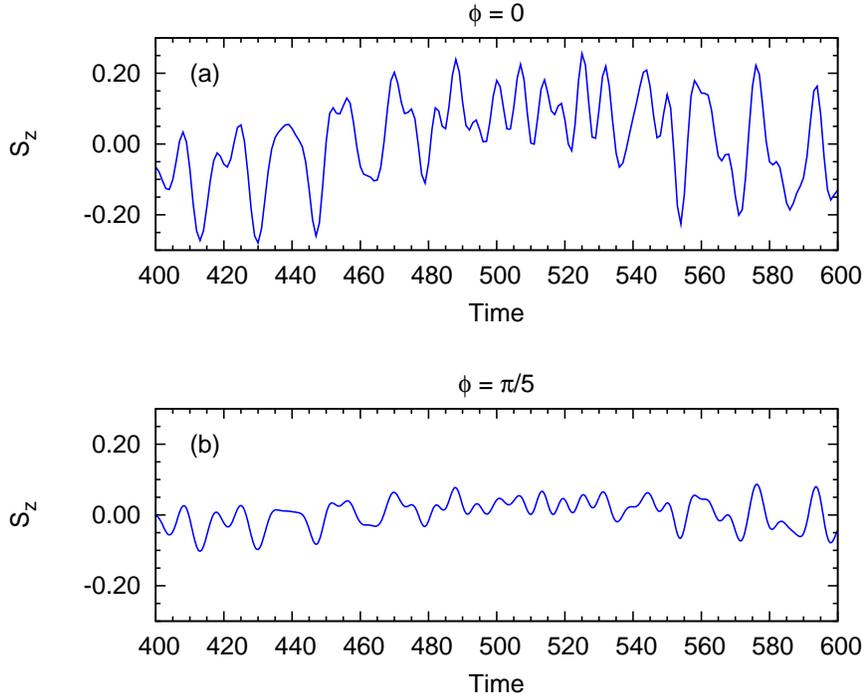}
\caption{The spin projection $S_z$ within a time subinterval.  The
fluctuations look like a series of split peaks.
The time is in units of $\hbar/V$
}
\label{Spinz}.
\end{figure}
\begin{figure}
\vspace{-0.8cm}
\centering
\includegraphics[width=8cm]{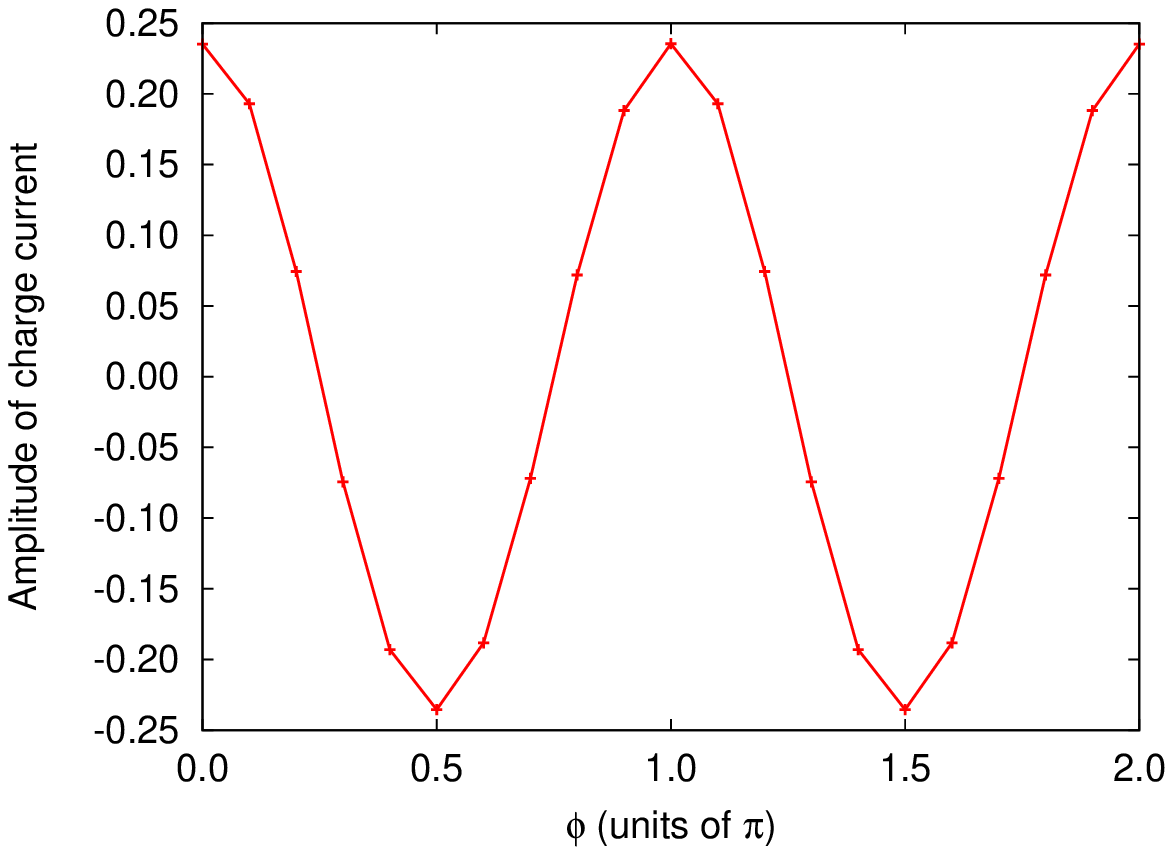}
\caption{ Amplitude of the charge current vs relative angle }
\label{amplit}.
\end{figure}

Next, in Fig.\ \ref{chargeRD}(b) we illustrate the time evolution of the
spin current, defined by $I_{S_z}=\frac{\hbar}{2} \left( \sigma_z v_{\theta}+ v_{\theta}\sigma_z \right) \,$, i.\ e.\ polarized along the $\hat{z}$ direction, for $\phi=0$. 
This instance indicates more clearly that the time evolution of
the system is not periodic even long after the external pulse has
vanished.  A striking beating pattern is observed, on a background of
fast oscillations, faster and much more pronounced than for the charge
current. To explain the fast oscillations we refer to Fig.\
\ref{Spinz}, which shows the time dependence of the spin projection on
the $z$ axis within a time interval of 200 units. A series
of double peaks is noticeable created by the two SOI couplings that generate spin
torque in opposite directions. The result is a superposition of Rashba and
Dresselhaus precessions, with a certain relative phase shift.  Consequently, the fast
oscillations seen in the charge current are doubled for $S_z$ and for the
the spin current.  The average time between consecutive maxima in Fig.\
\ref{Spinz} is about $7.4$ $\hbar/V$ which is about half of the time corresponding
to the fast oscillations of the charge current within the same time
interval.

We find, therefore, that the spin current is much more sensitive to
the single electron transitions between states with distant energies,
implicating high frequencies, than the charge current.  The beating
pattern is a result of that.  Since the  SOI parameters are different,
$V_{\alpha} \neq V_{\beta}$, the principal spin axes ${\bf e}_{\alpha}$
and ${\bf e}_{\beta}$ have different angles with the $z$ direction,
$\Omega_{\alpha}\neq \Omega_{\beta}$, and the electron spin executes
rapid oscillations between them.  In physical terms the orbital motion
is accompanied by spin nutation.

At equal SOI parameters, $V_{\alpha}=V_{\beta}$, the principal
directions do not coincide, but make the same angle with the $z$ axis,
$\Omega_\alpha=\Omega_\beta$.  The spin precessions cancel each other
and the corresponding spin current vanishes.  However the charge current
remains qualitatively as in Fig.\ \ref{chargeRD}.  If only one type of
SOI is present, the spin current is not zero, but constant, whereas the
charge current vanishes \cite{nita}, as shown by the green lines in Fig.\
\ref{chargeRD} (a) and (b).

In the lower panels of Figs.\ \ref{chargeRD}, (c) and (d), we show the
results for the charge and spin currents calculated at $\phi = \pi/5$.
When compared with the previous case, the amplitude of the slow
oscillations of the charge current decreases, while that of the spin
current increases.  Fast oscillations are still pronounced in the spin
current.  Moreover, the spin current shows similar oscillations as the
charge, with a phase shift of $\pi/2$, which dominate now the nutation
effects.  This effect can be explained on account of the anisotropy of
the charge distribution shown in Fig.\ \ref{CDW}.  The symmetry axis
of the ring corresponds to $\theta=\pi/4$, and therefore if
the pulse is oriented along this direction no charge current is induced.
This is seen further in Fig.\ \ref{amplit} where we show the amplitude of
the charge current vs. the angle $\phi$, which vanishes for $\phi=\pi/4$.
In this case a spin current of maximum amplitude is induced due to
the opposite chiralities of the states.  The slow oscillations overlap
with the fast fluctuations described earlier.  At $\phi=0$, the charge
current has a maximum amplitude, whereas that of the spin current is
minimal and reduced to the fast fluctuations related to spin nutation.
The angle $\phi=\pi/5$ is used in Fig.\
\ref{chargeRD}(c) and (d) as the closest value to $\pi/4$ which can be
obtained on our discretized ring with 20 sites.


\section{Conclusion}

The simultaneous effect of Rashba and Dresselhaus spin-orbit
interactions upon spin and charge currents are estimated numerically
in a ring geometry. A linearly polarized terahertz laser pulse is used
for a non-adiabatic generation of the currents. In the presence of
the dual spin-orbit coupling, the charge and spin currents exhibit a
quasi-periodic behavior whose amplitude depends on the relative angle
between the pulse and the charge deformation created by the SOI effects.
The spin current reflects a nutation of the electron spin between 
the Rashba and Dresselhaus spin quantization axes, superimposed as a 
fast beating pattern on the main oscillation.

\ack
This work was supported by the Icelandic Research Fund and the
Romanian PNCDI2 Research Programme and Core Programme (Contract No. 45N/2009).


\end{document}